\documentclass[11pt,a4paper]{article}
\usepackage[margin=1in]{geometry}
\usepackage[T1]{fontenc}
\usepackage[utf8]{inputenc}
\usepackage{lmodern}
\usepackage{microtype}
\usepackage{amsmath,amssymb,amsthm,mathtools}
\usepackage{booktabs,array,tabularx,longtable}
\usepackage[dvipsnames]{xcolor}
\usepackage{tikz}
\usetikzlibrary{arrows.meta,positioning,fit,shapes.geometric}
\usepackage{hyperref}
\hypersetup{colorlinks=true,linkcolor=MidnightBlue,citecolor=MidnightBlue,urlcolor=MidnightBlue,pdftitle={RummiAlgebra: A Polynomial Representation Framework for Game-Based Algebra Learning with Jokers},pdfauthor={Luis Adrian Lizama Perez},pdfkeywords={polynomial operations, RummiAlgebra, mathematics education, jokers}}
\usepackage[sort&compress,numbers]{natbib}
\usepackage{enumitem}
\usepackage{caption}
\setlist[itemize]{leftmargin=*,topsep=3pt,itemsep=2pt}
\setlist[enumerate]{leftmargin=*,topsep=3pt,itemsep=2pt}

\newtheorem{proposition}{Proposition}
\newtheorem{theorem}{Theorem}
\newtheorem{corollary}{Corollary}

\newcommand{\Tiles}{\mathcal{T}}

\newcommand{\colset}{\mathcal{C}}
\newcommand{\poly}{\Phi}
\newcommand{\score}{\operatorname{score}}

\title{\textbf{RummiAlgebra: A Polynomial Representation Framework for Game-Based Algebra Learning with Jokers}\\[4pt]
\large Mathematical Foundations, Correctness, Examples, and an Interactive Prototype}
\author{Luis Adri\'an Lizama P\'erez\\
\small Departamento de Sistemas de Informaci\'on y Comunicaciones\\[-1pt]
\small Universidad Aut\'onoma Metropolitana, Unidad Lerma, M\'exico\\[-1pt]
\small \texttt{l.lizama@ler.uam.mx}}
\date{November 2026 \quad (preprint draft, version 2.3)}
\begin{document}
\maketitle
\begin{abstract}
Connecting symbolic expressions to manipulable objects is a recurring challenge in introductory algebra. This paper presents \emph{RummiAlgebra}, a design-oriented framework and browser-based prototype that maps a Rummikub-style tile-rearrangement task to polynomial addition, common-factor extraction, finite geometric sums, and context-dependent substitution. A numbered colored tile is encoded as a monomial $c x^v$, while a physical joker is represented by a distinct symbol $J_i$ whose algebraic role is fixed only for a particular candidate arrangement. We define admissible polynomial forms for runs and groups and establish an inventory-conservation invariant. A soundness-and-completeness proof shows that physical-tile accounting together with the algebraic run/group templates characterizes valid final arrangements under the stated classroom rules; polynomial equality alone is insufficient. Four detailed examples illustrate decomposition, competing factorizations, one- and two-joker substitutions, and optimization as an optional extension. A targeted review distinguishes the approach from established Rummikub solvers, Rummikub-inspired arithmetic-sequence activities, algebra-tile manipulatives, and polynomial games. The prototype implements six small challenges and a separate rule-and-expression checker; their optimal scores were checked against an independent exhaustive reference enumerator. These checks establish agreement for the examples but do \emph{not} establish learning effectiveness or novelty of the underlying algebraic operations. We close with a classroom sequence, assessment constructs, and a controlled evaluation protocol. The claimed contribution is a formally specified pedagogical correspondence and an interactive proof-oriented activity, not a new asymptotically faster Rummikub solver.
\end{abstract}

\section{Introduction}
Students are often asked to expand or factor a polynomial without an accessible explanation of what has been preserved by the transformation. In a typical exercise, $x^3+x^4+x^5=x^3(1+x+x^2)$ is a symbolic identity; it is not immediately apparent why either side would be a more useful way of organizing the same information. Interactive manipulatives address this difficulty by connecting visual and symbolic forms, although the nature of the representations and the learning tasks matters \citep{Ainsworth2006,Cevikbas2021,Tondorf2022}.

A tile-rearrangement game offers a concrete reason to transform an expression. A player must preserve all tiles already on the table while dividing them into legal groups and runs. Different arrangements may use different numbers of tiles from the hand. The proposed \emph{RummiAlgebra} environment turns these actions into algebraic questions: What polynomial describes the tile inventory? What changes when tiles move from one set to another? Which common factors express the structure of a legal run or group? How should a joker be interpreted without treating it as an unrestricted algebraic equality?

This design is deliberately more specific than the broad claim that games can support mathematics learning. Board games, algebraic puzzles, and virtual algebra tiles already have substantial precedents \citep{Lee2023,Belloso2024,Garzon2018,Pereira2024}. Our focus is the conjunction of \emph{tile conservation, constrained polynomial forms, per-arrangement joker substitutions, and immediate bidirectional feedback}. The interface is intended to make a student's algebraic argument inspectable, not merely to display the answer supplied by an optimization engine.

The paper addresses three questions:
\begin{enumerate}
\item Can the rearrangement rules be connected to polynomial identities by a representation with a precise invariant?
\item Can jokers be incorporated as explicit, local substitutions without conflating a physical joker with a fixed monomial?
\item What part of the proposal is distinguishable from prior work, and what evidence is still needed before claiming an educational benefit?
\end{enumerate}

We answer the first two questions formally and by construction; Appendix~\ref{app:correctness} provides a compact correctness theorem for the rule-aware verification method. For the third, we offer a bounded literature review, a working prototype, and an evaluation plan. No participant study is reported.

\section{State of the Art and Positioning}
\label{sec:related}
\subsection{Review scope and search strategy}
The review is \emph{targeted and narrative}, not a systematic review or proof of global novelty. Searches were conducted in November 2026 on scholarly publisher pages, arXiv, institutional repositories, and education-journal records. Terms included combinations of \emph{Rummikub, Rummy, algebra, polynomial, factoring, algebra tiles, arithmetic progression, game-based learning, multiple representations, digital manipulatives,} and \emph{joker}. We prioritized sources offering an identifiable author, publication venue, year, and permanent link or DOI. We also checked nearby literature in Portuguese and Spanish, because a search restricted to English would miss pertinent teaching designs. Exact search replication, complete recall, and the absence of unpublished software cannot be guaranteed.

\subsection{Rummikub as a computational problem}
\citet{DenHertog2006} formulated maximizing the number or value of played tiles as an integer linear program and considered rearrangements of existing table sets. \citet{VanRijn2016} described a dynamic programming algorithm for runs and groups and analyzed the complexity of generalized instances, including the treatment of a bounded number of jokers. These are important precedents for \emph{computing} legal moves. Their main objective, however, is optimization rather than teaching the distributive property or demonstrating algebraic equivalence. We therefore do not present an exact solving algorithm as a novel contribution.

A rule specification is also essential: in the standard game a run has at least three consecutive numbers of one color; a group has three or four equal numbers in distinct colors; standard sets include two copies of each number-color tile and two jokers \citep{RummikubRules2025}. Physical steps and joker-recovery rules can depend on the adopted rule set. The classroom prototype consequently checks legal \emph{final configurations}, not every physical maneuver allowed by every edition.

\subsection{Rummikub and other board games in mathematics education}
Prior uses of Rummikub are closer to this paper's didactic motivation than the optimization literature. In the \emph{Acitemtirap} project, \citet{Soares2025} adapted the game to teach arithmetic progressions and reported classroom use and students' favorable perceptions. \citet{Lee2023} employed Rummikub along with other board games in a university mathematics course, discussing mathematical principles such as probability within a broader gamified teaching model. These works establish that bringing Rummikub into a mathematics classroom is not itself new. They do not, in the versions reviewed, define the specific colored-monomial, constrained-factorization, and joker-substitution calculus developed here.

The broader idea of teaching algebra through games also predates this work. \citet{Belloso2024} describes polynomial puzzles, factor-search activities, algebraic dominoes, and other board games. \citet{Rizzo2022} uses puzzles and riddles to teach factorization, while \citet{MarcoFigural2012} proposes a game-like figural approach to factoring quadratic polynomials. This literature is essential: it rules out characterizing a playful polynomial activity as inherently unprecedented.

\subsection{Digital manipulatives and representations of algebra}
\citet{Garzon2018} developed \emph{Virtual Algebra Tiles} and reported a study with 40 engineering students assigned to experimental or comparison conditions. Their reported learning and retention effects concern that particular tool and population; they must not be transferred to RummiAlgebra. \citet{Pereira2024} investigated polynomial operations using the Mathigon algebra-tiles resource with secondary students. \citet{Stemele2024} analyzed algebraic-expression errors and an algebra-tile intervention in Grade 9. These implementations emphasize geometric or area models and the link between manipulation and symbol use.

The DeFT framework of \citet{Ainsworth2006} explains how multiple external representations can complement, constrain, and construct understanding. In a task-specific study, \citet{Tondorf2022} distinguishes recognizing equivalent results from dynamically transforming one expression into another. Such findings motivate RummiAlgebra's simultaneous display of physical tiles, expanded terms, and factored groups. They also warn against assuming that simply displaying two representations ensures that students will coordinate them.

A systematic review of task design in dynamic mathematics environments \citep{Cevikbas2021} documents diverse study designs and practical challenges, while a 2025 scoping review of digital factorization tools \citep{Ortega2025} maps reported learning obstacles and technological interventions. A second-order meta-analysis \citep{Bozkus2026} examines aggregate game-based mathematics outcomes across existing reviews and their variation across contexts. None of these sources evaluates RummiAlgebra; accordingly, the prototype's existence cannot establish an educational benefit.

\subsection{Positioning relative to the closest precedents}
\begin{table}[t]
\centering\footnotesize
\caption{Closest approaches and the delimited scope of the present design. A ``not reported'' entry refers only to the cited source, not to a universal absence.}
\label{tab:position}
\begin{tabularx}{\textwidth}{@{}p{.24\textwidth}p{.29\textwidth}X@{}}
\toprule
\textbf{Approach} & \textbf{Established contribution} & \textbf{Distinction proposed here}\\
\midrule
Rummikub optimization \citep{DenHertog2006,VanRijn2016} & Legal-set optimization, dynamic programming, optional jokers & Learner-facing symbolic justifications; solver efficiency is not the claim\\
Rummikub in teaching \citep{Soares2025,Lee2023} & Arithmetic sequences, probability, game engagement & Polynomial inventory invariant and rule-constrained factorization\\
Algebraic games \citep{Belloso2024,Rizzo2022} & Polynomial puzzles, factor matching and games & Reconfiguration of the \emph{same} physical inventory with a simultaneous algebraic trace\\
Virtual algebra tiles \citep{Garzon2018,Pereira2024} & Visual polynomial manipulation and classroom studies & Game-specific run/group grammar; scoped joker assignments\\
Representation research \citep{Ainsworth2006,Tondorf2022} & Theory and evidence on coordinating representations & Concrete software operationalization and prospective assessment rubric\\
\bottomrule
\end{tabularx}
\end{table}

\noindent The defensible contribution is a \emph{design and formalization claim}: an integrated mapping between a particular tile game, constrained symbolic expressions, and explicit substitution-based feedback. Establishing priority against all existing software or demonstrating educational superiority would require substantially broader evidence.

\section{Mathematical Model}
\label{sec:model}
\subsection{Physical tiles and the polynomial inventory}
Let $\colset=\{r,b,g,k\}$ denote four \emph{formal indeterminates} for colors. The red, blue, green, and black symbols are labels; they are not numerical constants. A normal tile $t=(v,c)$ has number $v\in\{1,\dots,13\}$ and color $c\in\colset$. Multiple physical copies can have the same $(v,c)$ but carry distinct physical identifiers.

For a multiset $S$ of normal tiles, define the encoding
\begin{equation}\label{eq:phi}
\poly(S)=\sum_{(v,c)\in S} c\,x^v\in\mathbb{Z}[r,b,g,k,x].
\end{equation}
This is a \emph{polynomial encoding of the inventory}, not a numerical evaluation at a chosen $x$. If there are two red tiles numbered 6, their contribution is $2rx^6$. Colors must remain independent symbolic labels for the coefficients of distinct tile types to remain distinguishable.

\begin{proposition}[Faithfulness for normal-tile inventories]\label{prop:injective}
For normal-tile multisets $S,S'$, $\poly(S)=\poly(S')$ if and only if $S$ and $S'$ contain the same multiplicity of every number-color type.
\end{proposition}
\begin{proof}
The monomials $cx^v$ for distinct pairs $(c,v)$ are distinct monomials in $\mathbb{Z}[r,b,g,k,x]$. Equality of two polynomials requires equality of the integer coefficients of each monomial. Those coefficients are exactly the physical multiplicities of the associated tile type. The converse follows immediately by summation.
\end{proof}

The encoding supports elementary classroom operations. Adding a tile is polynomial addition; removing an available tile is subtraction; combining duplicate tile types produces a positive integer coefficient. Subtraction is legal as a tile action only when the corresponding multiplicity is positive. An unrestricted manipulation over $\mathbb{Z}$ might introduce negative coefficients that have no physical interpretation.

\subsection{Runs and groups as constrained factorizations}
A legal run of length $\ell\geq3$, color $c$, and starting number $a$ satisfies $1\leq a$ and $a+\ell-1\leq13$. Its encoding is
\begin{equation}\label{eq:run}
R_{c,a,\ell}(x)=c\sum_{i=0}^{\ell-1}x^{a+i}
=cx^a\underbrace{(1+x+\cdots+x^{\ell-1})}_{S_\ell(x)}.
\end{equation}
A legal group consists of $h\in\{3,4\}$ tiles of the same number $v$ and pairwise distinct colors $D\subseteq\colset$, with $|D|=h$:
\begin{equation}\label{eq:group}
G_{v,D}(x)=x^v\sum_{c\in D}c.
\end{equation}
The additional conditions on $\ell$, boundaries, coefficients, and distinct colors are part of these \emph{templates}. They do not follow from arbitrary factorability.

\begin{proposition}[Recognition of a single legal meld]\label{prop:recognition}
A multiset of normal tiles forms a legal run if and only if its encoding has the form in \eqref{eq:run} with each displayed power present exactly once and $\ell\geq3$. It forms a legal group if and only if its encoding has the form in \eqref{eq:group} with $|D|\in\{3,4\}$ and pairwise distinct colors.
\end{proposition}
\begin{proof}
The forward implications follow from the rules of a run or a group. Conversely, by Proposition~\ref{prop:injective}, each prescribed monomial identifies exactly one required tile type with multiplicity one. The run template gives consecutive admissible numbers of a single color, and the group template gives one number in distinct colors. These are precisely the respective legality conditions.
\end{proof}

An especially useful identity arises from the finite geometric series:
\begin{equation}\label{eq:telescoping}
(1-x)S_\ell(x)=1-x^\ell,
\qquad
(1-x)\,R_{c,a,\ell}(x)=cx^a-cx^{a+\ell}.
\end{equation}
It follows by distributivity and cancellation:
$\sum_{i=0}^{\ell-1}x^i-\sum_{i=1}^{\ell}x^i=1-x^\ell$.
The pair of endpoint terms is a concise certificate \emph{when the proposed run already satisfies the support and coefficient restrictions}. For a sum of overlapping runs, the difference can contain other signed endpoint contributions. The identity alone does not reconstruct a unique decomposition into runs.

\subsection{Jokers as scoped substitutions}
Physical jokers must remain distinguishable from ordinary tiles. If $J_1,\dots,J_q$ denote their unique identifiers, define
\begin{equation}\label{eq:physical}
P_{\rm phys}(S)=\poly(S_{\rm normal})+\sum_{i\in I(S)}J_i
\in\mathbb{Z}[r,b,g,k,x,J_1,\ldots,J_q].
\end{equation}
A candidate arrangement $A$ assigns each \emph{played} joker a role $(v_i,c_i)$. This defines a ring homomorphism $\sigma_A$ by
\begin{equation}\label{eq:sigma}
\sigma_A(J_i)=c_i x^{v_i},\qquad
\sigma_A(x)=x,\quad \sigma_A(c)=c \;(c\in\colset),
\end{equation}
with unplayed jokers left as $J_i$. The substitution is contextual: $\sigma_A(J_i)=rx^6$ for one arrangement does not assert the identity $J_i=rx^6$ in the physical inventory ring, and another arrangement may assign the same physical joker a different role.

Each joker occupies exactly one position within one proposed meld. After substitution, the resulting \emph{effective} run or group must satisfy the constraints in Proposition~\ref{prop:recognition}; in particular, two symbols with equal colors within a three-color group are not permitted merely because the expanded polynomials can be added. When two jokers are present, their assignments are jointly constrained by the proposed melds. The model permits distinct candidate assignments to be evaluated and compared.

\subsection{Conservation, legality, and objective}
Let $\Tiles=\Tiles_{\rm table}\sqcup\Tiles_{\rm hand}$ denote the physically indexed inventory. An arrangement $A$ consists of disjoint sets $M_1,\dots,M_s$ and unused tiles $U$ such that
\begin{equation}\label{eq:partition}
\Tiles=U\sqcup M_1\sqcup\cdots\sqcup M_s.
\end{equation}
For a classroom task with an existing table, we additionally require $\Tiles_{\rm table}\cap U=\varnothing$: all original table tiles remain in some valid meld. A candidate that leaves a partial run or group on the table is rejected. After applying $\sigma_A$ to both sides of the physical partition, we obtain
\begin{equation}\label{eq:invariant}
\sigma_A\bigl(P_{\rm phys}(\Tiles)\bigr)
=\sigma_A\bigl(P_{\rm phys}(U)\bigr)
+\sum_{j=1}^s \sigma_A\bigl(P_{\rm phys}(M_j)\bigr).
\end{equation}

\begin{proposition}[Conservation is necessary but not sufficient]\label{prop:necessary}
Every legal arrangement satisfies \eqref{eq:invariant}. The converse is false unless physical identity, partition conditions, per-meld legality, and any table or opening constraints are checked separately.
\end{proposition}
\begin{proof}
Every legal arrangement is a disjoint partition of its physical tile inventory. Polynomial encoding and substitution preserve addition, establishing the equality. Conversely, place $3R$, $5R$, and $7R$ in one purported run. The expression $rx^3+rx^5+rx^7=rx^3(1+x^2+x^4)$ is a valid factorization and respects tile conservation, but the powers are not consecutive. Thus the set is illegal despite polynomial equality. Similar counterexamples arise by repeating a color in a group.
\end{proof}

For a played tile $t$ from the hand, define its score as its number $v$ if normal, or $v_i$ if it is a joker with assigned role $(v_i,c_i)$. The \emph{teaching instance} may optionally ask the student to maximize
\begin{equation}\label{eq:score}
\score(A)=\sum_{t\in\Tiles_{\rm hand}\setminus U}v_{\sigma_A}(t).
\end{equation}
This serves to compare valid decompositions, not to define algebraic correctness. The browser prototype checks the final partition and uses the assigned value to score a hand joker. It does not implement every official joker-recovery procedure.

A proposed arrangement is accepted only when its physical tile identifiers form a disjoint partition, its played jokers have explicit roles, each effective meld matches one of the constrained polynomial templates, and the applicable table/opening conditions hold. Appendix~\ref{app:correctness} proves that these checks are both necessary and sufficient for legality under the educational model. The same appendix explains when exhaustive enumeration is guaranteed to find an optimum.

\section{From Game Moves to Algebraic Operations}
\label{sec:pedagogy}
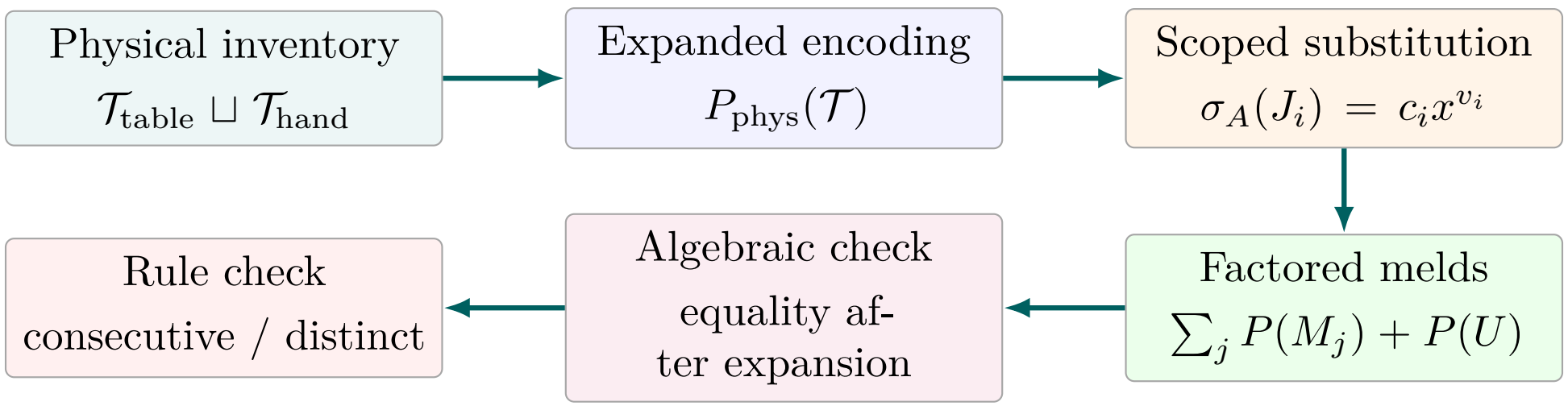
\begin{figure}[t]
\centering
\begin{tikzpicture}[node distance=7mm and 10mm, >=Latex,
nodebase/.style={draw=black!35,rounded corners=2pt,align=center,minimum height=9mm,text width=33mm,font=\small,inner sep=4pt},
arr/.style={-{Latex[length=2.2mm]},very thick,draw=teal!75!black}]
\node[nodebase,fill=teal!7] (p) {Physical inventory\\\smallskip $\Tiles_{\rm table}\sqcup\Tiles_{\rm hand}$};
\node[nodebase,right=of p,fill=blue!5] (e) {Expanded encoding\\\smallskip $P_{\rm phys}(\Tiles)$};
\node[nodebase,right=of e,fill=orange!9] (s) {Scoped substitution\\\smallskip $\sigma_A(J_i)=c_ix^{v_i}$};
\node[nodebase,below=of s,fill=green!8] (f) {Factored melds\\\smallskip $\sum_j P(M_j)+P(U)$};
\node[nodebase,left=of f,fill=purple!7] (c) {Algebraic check\\\smallskip equality after expansion};
\node[nodebase,left=of c,fill=red!6] (v) {Rule check\\\smallskip consecutive / distinct};
\draw[arr] (p)--(e);\draw[arr](e)--(s);\draw[arr](s)--(f);\draw[arr](f)--(c);\draw[arr](c)--(v);
\end{tikzpicture}
\caption{Physical, symbolic, and rule-based checks are linked but not interchangeable. For cases without jokers, the substitution stage is the identity.}
\label{fig:levels}
\end{figure}

Figure~\ref{fig:levels} shows the intended representational sequence. It makes the mathematical dependency explicit: the student moves physical tiles and writes a symbolic inventory, proposes a particular factorization, applies any joker substitutions, and checks both algebraic and game constraints.

The activity can present five algebraic operations in a coherent progression:
\begin{enumerate}
\item \textbf{Encoding and addition.} Convert a hand and an existing table to a polynomial; recognize coefficients as duplicate counts.
\item \textbf{Removal and conservation.} Split the expression into played and unplayed tiles. Explain why a missing or duplicated monomial reveals an inventory error.
\item \textbf{Distributivity and common factors.} Write a run as $cx^a(1+x+\cdots)$ and a group as $x^v\sum c$. Expand to confirm equality.
\item \textbf{Finite geometric sums.} Derive \eqref{eq:telescoping} and use cancellation to describe run endpoints.
\item \textbf{Substitution.} Assign a joker a role in one arrangement, and explain why another arrangement may require a different assignment.
\end{enumerate}

This sequence is \emph{not} a claim that the game supplies the full school-algebra curriculum. It especially does not teach arbitrary polynomial long division, roots, or quadratic factorization without additional activities. The particular mathematical scope is addition, factoring of constrained expressions, finite geometric sums, and substitution.

\section{Worked Examples}
\label{sec:examples}
\subsection{Rearranging a run to form a group}
\label{sec:example-main}
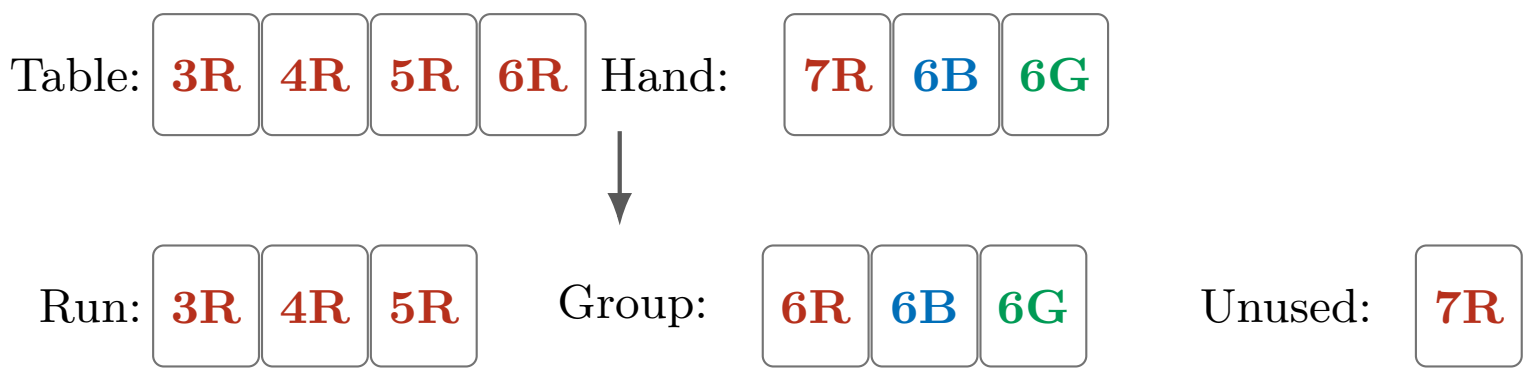
\begin{figure}[t]
\centering
\begin{tikzpicture}[x=0.72cm,y=0.85cm,
 tile/.style={draw=black!55,fill=white,rounded corners=2pt,minimum width=7mm,minimum height=8mm,font=\footnotesize\bfseries,inner sep=3pt},
 lab/.style={font=\footnotesize,anchor=east}]
\node[lab] at (-.4,1.35) {Table:};
\foreach \v/\p in {3/0,4/1,5/2,6/3}{\node[tile,text=BrickRed] at (\p,1.35) {\v R};}
\node[lab] at (5.0,1.35) {Hand:};
\node[tile,text=BrickRed] at (5.8,1.35) {7R};
\node[tile,text=NavyBlue] at (6.8,1.35) {6B};
\node[tile,text=ForestGreen] at (7.8,1.35) {6G};
\draw[-{Latex[length=2.2mm]},thick,black!65] (3.8,.91)--(3.8,.17);
\node[lab] at (-.4,-.45) {Run:};
\foreach \v/\p in {3/0,4/1,5/2}{\node[tile,text=BrickRed] at (\p,-.45) {\v R};}
\node[lab] at (4.8,-.45) {Group:};
\node[tile,text=BrickRed] at (5.6,-.45) {6R};
\node[tile,text=NavyBlue] at (6.6,-.45) {6B};
\node[tile,text=ForestGreen] at (7.6,-.45) {6G};
\node[lab] at (10.9,-.45) {Unused:};
\node[tile,text=BrickRed] at (11.6,-.45) {7R};
\end{tikzpicture}
\caption{The same seven physical tiles before and after a valid rearrangement. The proposed move plays both number-6 tiles from the hand and leaves the hand's red 7 unused.}
\label{fig:rearrange}
\end{figure}

Figure~\ref{fig:rearrange} depicts the before/after move. The abbreviated color letters inside each tile make the figure readable without color printing.
The table holds four red tiles $3R,4R,5R,6R$, and the player's hand contains $7R,6B,6G$. The initial inventory is
\begin{equation}\label{eq:exampleoriginal}
P_{\rm phys}=r(x^3+x^4+x^5+x^6+x^7)+bx^6+gx^6.
\end{equation}
One valid move extends the red run to $3R$--$7R$ and leaves both other tiles unused:
\[
P_{\rm phys}=rx^3(1+x+x^2+x^3+x^4)+bx^6+gx^6.
\]
Its score from the hand is $7$. A better move partitions the same normal-tile inventory as
\begin{align}\label{eq:examplebest}
P_{\rm phys}&=r(x^3+x^4+x^5)+x^6(r+b+g)+rx^7\\
&=rx^3(1+x+x^2)+x^6(r+b+g)+rx^7.\nonumber
\end{align}
The first term represents the legal red run $3R$--$5R$, the second the legal group $6R,6B,6G$, and the final term the unused hand tile $7R$. Every original table tile is included. The hand contributes $6+6=12$ points. This is optimal for the seven-tile instance; the reference enumerator in Section~\ref{sec:implementation} confirms the value.

This example explains the algebraic meaning of ``the same tiles, a different factorization.'' Crucially, the groups themselves are identified by constrained patterns, rather than by arbitrary factorization of the entire polynomial.

\subsection{One joker bridges a run}
Consider the hand $5R,7R,J_1$. In the physical ring,
\[
P_{\rm phys}=rx^5+rx^7+J_1.
\]
We propose $\sigma_A(J_1)=rx^6$, which yields
\begin{align*}
\sigma_A(P_{\rm phys})&=rx^5+rx^6+rx^7\\
&=rx^5(1+x+x^2).
\end{align*}
The sequence is consecutive, uses one color, and contains three physical tiles. The joker's assigned value is $6$, so the played-hand value in this pedagogical scoring model is $5+6+7=18$. By contrast, $\sigma_A(J_1)=bx^6$ gives $rx^5+bx^6+rx^7$, which is not a single-color run. The algebraic sum exists but does not validate that meld.

\subsection{A joker completes a group}
Given $9R,9B,J_1$, choose $\sigma_A(J_1)=gx^9$. Then
\[
rx^9+bx^9+gx^9=x^9(r+b+g).
\]
The three colors are distinct and the score is $9+9+9=27$. A black joker assignment would also be locally valid. An assignment of $J_1$ to a \emph{second red} $9$ produces a factorization $x^9(2r+b)$, but is not a legal group because red occurs twice. The difference between an identity and a legal grouping is now visible to the student.

\subsection{Two jokers and competing substitutions}
The hand consists of $4R,7R,J_1,J_2$. One arrangement uses all four tiles in a red run:
\begin{align*}
\sigma_A(J_1)=rx^5,\quad\sigma_A(J_2)=rx^6,
&\qquad\sigma_A(P_{\rm phys})=rx^4(1+x+x^2+x^3),\\
&\score(A)=4+5+6+7=22.
\end{align*}
Another arrangement leaves $4R$ unused and forms $7R,8R,9R$:
\begin{align*}
\sigma_B(J_1)=rx^8,\quad \sigma_B(J_2)=rx^9,
&\qquad \sigma_B(P_{\rm phys})=rx^4+rx^7(1+x+x^2),\\
&\score(B)=7+8+9=24.
\end{align*}
Both are legal. The second has the higher score in the stated objective, but the two expressions use \emph{different substitutions}; their equality should never be asserted without the $\sigma_A$ or $\sigma_B$ qualification. This exercise differentiates \emph{variable meaning}, \emph{equality after substitution}, and \emph{optimization}.

\section{Interactive Prototype and Reproducible Checks}
\label{sec:implementation}
\subsection{Prototype features}
The accompanying prototype, \emph{RummiAlgebra V2}, is a standalone HTML/JavaScript file that opens locally in a modern browser. It offers a physical tile board, an algebra panel, six guided challenges, and a custom exercise editor. The learner may move tiles by selection or drag-and-drop, select a run or group, assign each played joker a number and color, request a compatible suggestion, apply the relevant factor extraction, inspect the expanded identity, and check the final arrangement. The algebra panel distinguishes the physical inventory, the scoped joker substitution, the substituted inventory, and the rearranged expression.

The checker accepts a completed challenge only after inspecting the physical tile partition, every run or group, the assignments of played jokers, the table-preservation rule, and the presence of appropriate factorization justifications. A separate small-instance optimizer can show a reference arrangement, but it is optional to the learning task. The prototype limits automatic solutions to at most 15 tiles without jokers or 14 with jokers and permits manual checking up to 30 tiles; these are implementation limits, not mathematical bounds. The interface does not log educational outcomes or send user data to a server.

\subsection{Verification scope}
We checked the six packaged puzzles with an \emph{independent exhaustive reference enumerator}, implemented in Python without importing the browser solver. It enumerates candidate physical-tile subsets, legal runs and groups, and permissible one- or two-joker assignments. A subset-selection search then finds the maximum sum of the played-hand values subject to mandatory table-tile use. This is an independent cross-check of the listed small examples, not a formal verification of every browser event handler.

\begin{table}[htbp]
\centering\small
\caption{Six built-in examples: published teaching target and independently enumerated optimum. ``Opening'' includes the 30-point initial-meld condition.}
\label{tab:tests}
\begin{tabular}{@{}clcc@{}}\toprule
\textbf{ID} & \textbf{Activity} & \textbf{Prototype target} & \textbf{Reference optimum}\\\midrule
1 & Table rearrangement & 12 & 12\\
2 & Opening (30-point threshold) & 30 & 30\\
3 & Regrouping existing tiles & 42 & 42\\
4 & One joker in a run & 18 & 18\\
5 & One joker in a group & 27 & 27\\
6 & Two-joker comparison & 24 & 24\\\bottomrule
\end{tabular}
\end{table}

The exhaustive reference reports $6/6$ agreements. This is a useful implementation check, but it cannot establish general correctness of a large solver or, more importantly, improvement in student understanding. The verifier and the six exact inputs are included in the reproducibility package. The browser prototype and the independent verifier operate on the stated educational rules, not all tournament-specific conventions.

\subsection{A minimal conceptual checking procedure}
A rule-based check is much easier to explain to a beginner than a global optimum solver. The following pseudocode outlines the sequence:
\begin{quote}
\small\ttfamily
INPUT: indexed physical tiles; proposed groups/runs; joker roles.\\
CHECK: each tile appears once; no original table tile is unused.\\
FOR each proposed meld:\\
\hspace*{1em}substitute any played jokers by their chosen roles;\\
\hspace*{1em}check one-color consecutive run OR equal-value distinct-color group;\\
\hspace*{1em}derive its allowed factored form and expand it.\\
CHECK: sum of expanded melds and unused tiles equals substituted inventory.\\
OUTPUT: legal/illegal with explanation; optional hand-tile score.
\end{quote}

The order matters. Polynomial equality is a conservation check, whereas the run/group grammar is a structural check. A learner benefits from knowing \emph{which} condition failed.

\section{Instructional Design and Assessment Plan}
\label{sec:evaluation}
\subsection{A classroom learning sequence}
A proposed 55-minute activity is organized around explainable transformations rather than time pressure:
\begin{enumerate}
\item \textbf{Inventories (10 min):} encode five to seven tiles; identify colors as symbolic labels, values as exponents, and duplicate copies as coefficients.
\item \textbf{Representations (10 min):} expand and factor one legal run and one group; ask students to predict the tile arrangement before looking at feedback.
\item \textbf{Rearrangement (15 min):} solve Example~\ref{sec:example-main}; require a written argument showing both preservation and legality.
\item \textbf{Jokers (12 min):} assign one and then two joker roles; contrast an algebraically well-formed sum with an illegal proposed meld.
\item \textbf{Reflection (8 min):} explain why two distinct factorizations can describe the same inventory and why an arbitrary factorization need not be a legal move.
\end{enumerate}

The teacher should ask students to interpret $r$ as an \emph{indeterminate naming a color}, not as ``the number of red tiles.'' The latter is encoded by a coefficient. In addition, the joker assignment must be explicitly indexed by the arrangement; otherwise students may believe that $J_1=rx^6$ is a permanent identity.

\subsection{Observable learning constructs}
A future study should assess specific outcomes rather than treating ``enjoyment'' as synonymous with conceptual learning. A four-part rubric could use 0--2 points for each construct:
\begin{enumerate}
\item \emph{Encoding accuracy}: correct numbers, colors, multiplicities, and distinction between physical joker and assigned role.
\item \emph{Equivalent transformation}: correct distributivity and common-factor extraction, with all terms preserved.
\item \emph{Rule reasoning}: correct differentiation between legal melds and merely equivalent expressions.
\item \emph{Transfer}: successful solution of an equivalent task without physical tiles or with unfamiliar symbolic labels.
\end{enumerate}
The total score would range from 0 to 8. This rubric is a \emph{proposal}, not an already validated measurement instrument. Think-aloud or written explanations should accompany correctness scores, particularly for the substitution tasks.

\subsection{Prospective empirical design}
A controlled evaluation could randomly allocate students or intact classes to (i) RummiAlgebra-supported instruction or (ii) equal-duration instruction with conventional algebra tasks and equivalent teacher feedback. Both conditions should cover the same target identities, receive the same number of practice opportunities, and use parallel pretest, immediate posttest, and delayed posttest forms. The primary endpoint would be a pre-registered gain in algebraic equivalence and factorization accuracy on unfamiliar symbolic problems. Secondary endpoints could include rule-versus-identity discrimination, explanation quality, and engagement. Analyses should adjust for baseline proficiency and, if intact classes are assigned, account for clustering.

A pilot should first establish interface usability, item difficulty, scoring consistency, and potential ceiling effects. A subsequent sample-size calculation must use a justified effect size and intraclass correlation rather than selecting a convenient number of participants. Because the intervention includes game actions, feedback, and extra representational support, an additional dismantling comparison (with and without the synchronized algebra panel) would be needed to attribute learning gains to the polynomial correspondence itself.

\section{Limitations and Threats to Validity}
\label{sec:limits}
\textbf{Scope of algebra.} Factoring legal runs and groups demonstrates constrained common-factor extraction, not arbitrary algebraic factorization. Long division appears only in the special finite-geometric-series identity. If the educational objective is general polynomials, further tasks are required.

\textbf{Representation risk.} Color labels represented by indeterminates may increase cognitive load for novice students, and the analogy between rearrangement and factorization can be overextended. The requirement that legality be checked \emph{separately} is not optional.

\textbf{Joker convention.} A per-arrangement joker substitution is mathematically precise, but actual rule books may impose additional conditions on reclaiming or manipulating a joker already on the table. The present checker verifies final melds and mandatory table conservation, not a universally valid physical move sequence.

\textbf{Prototype scope.} The six reference cases are small and purpose-built. The manual checker is not a full-game strategy engine, and the automatic optimization mode is size-limited. No student data, learning-gain estimates, or effect sizes have been collected for RummiAlgebra.

\textbf{Prior-art scope.} The state-of-the-art search identifies strong overlapping precedents. The originality claim is limited to the stated combination of representations, substitution semantics, and rule-aware feedback as documented in the present prototype; it is not a claim of first use worldwide.

\textbf{Licensing and naming.} Rummikub is a proprietary game name. The prototype is a pedagogical adaptation and is not represented as an official product or an authorized endorsement. Reuse of game artwork, trademarks, or commercial branding requires separate review.

\section{Conclusion}
RummiAlgebra provides a precisely defined bridge between physical tile rearrangement and polynomial operations. Distinct monomials encode number-color tile types, run and group templates expose common factors, finite geometric sums make consecutive powers structurally visible, and scoped homomorphisms model joker assignments without treating them as permanent identities. The inventory-conservation invariant is exact, but it intentionally does not replace a legal-meld checker. A standalone browser prototype implements these ideas, and six illustrative instances match an independent small-instance enumerator. Appendix~\ref{app:correctness} proves soundness and completeness of the mathematical checking procedure for final arrangements, and shows why an exhaustive finite search can optimize its declared objective. These are model-level guarantees, not formal verification of the browser implementation. The primary research contribution is a formal educational design with explicit conceptual boundaries. Whether it improves algebra learning is an empirical question reserved for a controlled classroom study.

\paragraph{Software availability and reproducibility.}
The standalone RummiAlgebra V2 prototype can be used without installation at
\url{https://luislizama01.github.io/rummialgebra/}.
The source code, six classroom tasks, instructions, and independent small-instance
reference checker are available at
\url{https://github.com/luislizama01/rummialgebra}.
The tested software release is \texttt{v2.0.0}, available at
\url{https://github.com/luislizama01/rummialgebra/releases/tag/v2.0.0}
and archived by Zenodo under the version-specific DOI
\href{https://doi.org/10.5281/zenodo.24323391}{\texttt{10.5281/zenodo.24323391}} \citep{RummiAlgebraSoftware2026}.
The reference checker can be run with \texttt{python tests/verify\_examples.py};
its six expected scores were reproduced in our local verification.
The GitHub Pages site follows the mutable \texttt{main} branch, whereas the DOI
identifies the archived release; reproductions of this study should use the
latter. No external student data were used.

\appendix
\section{Correctness of the Algebraic Verification Method}
\label{app:correctness}

This appendix establishes correctness for the \emph{final-arrangement model} defined in Section~3. It does not address the efficiency of finding an arrangement or rule variations governing the physical recovery of a joker already on the table.

\paragraph{Certificate and verification.}
Let $\Tiles$ be a finite collection of \emph{individually identified} physical tiles, divided into table tiles $\Tiles_{\rm table}$ and hand tiles $\Tiles_{\rm hand}$. A proposed solution (certificate) specifies unused tiles $U$, nonempty melds $M_1,\ldots,M_s$, and a role $\rho(J_i)=(v_i,c_i)$ for every joker placed in a meld. The verifier performs the following checks:
\begin{enumerate}[label=\textup{(C\arabic*)},leftmargin=2.5em]
\item \textbf{Physical accounting:} the lists $U,M_1,\ldots,M_s$ contain each physical tile identifier exactly once; in particular, $\Tiles=U\sqcup M_1\sqcup\cdots\sqcup M_s$.
\item \textbf{Joker roles:} each played joker has one assignment $(v_i,c_i)\in\{1,\ldots,13\}\times\colset$, used consistently in its meld. Unused jokers keep their physical symbols.
\item \textbf{Algebraic meld forms:} after applying $\sigma_A$, each $M_j$ has \emph{exactly} one of the following polynomials:
\begin{equation}\label{eq:appendix-templates}
 cx^a\sum_{i=0}^{\ell-1}x^i\quad
 (\ell\ge3,\ 1\le a\le14-\ell),
 \qquad\text{or}\qquad
 x^v\sum_{c\in D}c\quad (|D|\in\{3,4\}),
\end{equation}
where $c\in\colset$, $1\le v\le13$, and $D\subseteq\colset$ has distinct elements. Equality means equality of formal polynomials, including coefficients; no extra terms are allowed.
\item \textbf{Task conditions:} no original table tile belongs to $U$; if the task requires an actual turn, at least one hand tile is used; and any explicitly imposed opening threshold is met.
\end{enumerate}
The verifier also checks the expanded inventory identity~\eqref{eq:invariant} as an explanatory algebraic trace. Under (C1)--(C2), this identity already follows from additivity; it is not a substitute for (C1) or (C3).

\begin{theorem}[Soundness and completeness]\label{thm:correctness}
Under the final-arrangement rules specified in Section~3, a proposed solution is accepted by (C1)--(C4) \emph{if and only if} it is a legal final arrangement for that task.
\end{theorem}
\begin{proof}
\emph{Soundness.} Suppose the checks pass. Condition (C1) prevents both loss and reuse of a physical tile. By (C2), each played joker contributes exactly one specified number--color role. Condition (C3), together with Proposition~\ref{prop:recognition}, makes every effective meld either a one-color run with consecutive distinct numbers and length at least three, or a same-number group with three or four distinct colors. Hence every meld is legal. Condition (C4) preserves the original table, ensures a nonempty play when required, and enforces any opening threshold. Therefore the complete arrangement is legal.

\emph{Completeness.} Conversely, take any legal final arrangement. List its actual unused tiles and melds, retaining the identifier of every physical tile; this satisfies (C1). Give each played joker the role that it occupies in that arrangement, satisfying (C2). Every legal run or group has, by definition, exactly the corresponding form in~\eqref{eq:appendix-templates}, so (C3) holds. The table, hand-play, and opening requirements of a legal task establish (C4). Therefore the verifier accepts.\qedhere
\end{proof}

\paragraph{Why polynomial equality alone cannot prove correctness.}
The formal identity $rx^3+rx^5+rx^7=rx^3(1+x^2+x^4)$ preserves all terms but describes an illegal run. Also, if two physical copies $t_1,t_2$ are both $6R$, the polynomial $2rx^6$ cannot distinguish using $t_1$ twice from using each copy once: condition (C1) is therefore indispensable. For a legal joker example, $\{5R,J_1,7R\}$ with $\sigma_A(J_1)=rx^6$ yields $rx^5(1+x+x^2)$; assigning $rx^8$ instead fails (C3).

\begin{corollary}[Correctness and optimality of exhaustive search]\label{cor:optimal}
Suppose a search enumerates every partition of the finite set of identified tiles into unused tiles and candidate melds, and every assignment of a permitted role to each played joker. It retains exactly the certificates accepted by (C1)--(C4). Then it (i) reports infeasibility if and only if no legal final arrangement exists; otherwise (ii) returns a legal arrangement maximizing the declared score~\eqref{eq:score}.
\end{corollary}
\begin{proof}
The search terminates because there are finitely many partitions of $\Tiles$ and at most $52$ permitted roles per played joker. By Theorem~\ref{thm:correctness}, the accepted certificates represent precisely the legal final arrangements. Taking the maximum score over this finite set therefore returns the global optimum, or correctly reports an empty feasible set.\qedhere
\end{proof}

\paragraph{Scope of the guarantee.}
The theorem certifies the \emph{mathematical specification} of the checker, not an arbitrary software implementation. For example, a UI that fails to enforce (C1) or overlooks a joker assignment does not inherit the theorem. Similarly, the corollary requires a genuinely exhaustive generator; the interactive exercises and their six reference tests are useful checks but do not constitute a machine-checked proof of the HTML code.

\section{Additional Algebraic Checks}

\subsection{Conservation with a played joker}
For normal tiles $5R,7R$ and joker $J_1$, the physical inventory is $rx^5+rx^7+J_1$. Applying the arrangement-specific homomorphism $\sigma_A(J_1)=rx^6$ yields $rx^5+rx^6+rx^7$. This polynomial equals $rx^5(1+x+x^2)$ by distributivity. It does \emph{not} equal the physical polynomial as a formal element of $\mathbb Z[r,b,g,k,J_1,x]$: the two expressions live on opposite sides of the substitution. Writing $P_{\rm phys}=rx^5(1+x+x^2)$ without mentioning $\sigma_A$ would be mathematically incorrect.

\subsection{Two copies are coefficients, not new exponents}
If a player holds two copies of $6R$ and one $7R$, the polynomial is $2rx^6+rx^7$, not $rx^{12}+rx^7$. The former preserves tile multiplicities while the latter changes the numerical labels. Such a distinction is necessary for the uniqueness result in Proposition~\ref{prop:injective}.

\subsection{The non-converse of the telescoping heuristic}
For any polynomial $P(x)$, $(1-x)P(x)$ is defined. The mere existence of a short or factorizable product does not imply that the support of $P$ consists of a single consecutive run. The recognition theorem therefore checks the pattern of nonzero monomials and their unit coefficients explicitly. For example, $P(x)=rx^3+rx^5+rx^7$ satisfies a common-factor identity but is not a run.

\section{Search and Evidence Record}
This draft's targeted review was audited against publisher records, arXiv, academic repositories, and original game-rule documentation in November 2026. Bibliographic metadata were compared at the level of authorship, year, journal or venue, DOI or stable URL, and relevance of each source to its cited statement. The compared sources are organized into \emph{computational Rummikub} \citep{DenHertog2006,VanRijn2016}, \emph{Rummikub classroom adaptations} \citep{Soares2025,Lee2023}, \emph{algebraic games} \citep{Belloso2024,Rizzo2022,MarcoFigural2012}, \emph{digital polynomial manipulatives} \citep{Garzon2018,Pereira2024,Stemele2024}, \emph{representational and review literature} \citep{Ainsworth2006,Tondorf2022,Cevikbas2021,Ortega2025,Bozkus2026}, and the \emph{versioned software artifact} \citep{RummiAlgebraSoftware2026}. A systematic priority claim or a claim about all commercially available education software would require a substantially broader and reproducible database search.

\bibliographystyle{plainnat}
\bibliography{references}
\end{document}